\documentclass{ws-ijmpcs}
\include{Qcircuit}

\begin{document}

\markboth{Y.\ Kondo \& M.\ Matsuzaki}
{Study of Open Systems with Molecules in Isotropic Liquids}

%
\catchline{}{}{}{}{}
%

\title{Study of Open Systems with Molecules in Isotropic Liquids}

\author{Yasushi Kondo}

\address{Department of Physics \& Science and 
Technology Research Institute, Kindai University, Higashi-Osaka, 
Osaka 577-8502, Japan\\
ykondo@kindai.ac.jp}

\author{Masayuki Matsuzaki}

\address{Department of Physics, Fukuoka University of Education, \\
Munakata, Fukuoka 811-4192, Japan
\\
matsuza@fukuoka-edu.ac.jp}

\maketitle

\begin{history}
\published{Day Month Year}
\end{history}

\begin{abstract}

We are interested in dynamics of a system in an environment, or an open system.
Such phenomena like {\it crossover} from Markovian to non-Markovian relaxation 
and  {\it thermal equilibration} are of our interest. 
Open systems have experimentally been studied with ultra cold atoms, 
ions in traps, optics, and cold electric circuits 
because well isolated systems can be prepared here 
and thus the effects of environments can be controlled. We point out that 
some molecules solved in isotropic liquid are well isolated 
and thus they can also be employed for 
studying open systems in 
Nuclear Magnetic Resonance (NMR) experiments.  
First, we provide a short review on related phenomena
of open systems  that helps readers to understand our motivation. 
We, then, present two experiments as examples of our approach with 
molecules in isotropic liquids. Crossover from Markovian to non-Markovian 
relaxation was realized in one NMR experiment, while relaxation like phenomena 
were observed in approximately isolated systems in the other. 

\end{abstract}

\section{Introduction}
\label{sec_intro}

A glass of cold beer becomes warm on a hot summer day. 
It is because the glass of 
beer interacts with its environment (often called a bath or lattice in some
fields) and it is not 
an isolated system.
It is an open system in the sense that it interacts with its environment. 
All the systems, except for the Universe itself, 
are not in principle isolated. Therefore, it is very important 
to understand an open system. 

An open system has been 
studied as a problem of thermodynamics 
and statistical mechanics from Maxwell's age. However,
most of studies were and are theoretical and experimental ones have only 
recently emerged. For example, a realization of {\it Maxwell's demon}
was first reported as late as in 2010\cite{Toyabe:2010aa}. 
This may be caused by the difficulty of controlling an environment: unknown 
factors are often attributed to the environment. In order to obtain a well 
controlled environment, it is necessary to have a well isolated system and then 
to increase the interaction between the system and its environment in a
controlled manner. Therefore,  we now 
expect the advance of 
this field by experiments with ultra cold atoms\cite{RevModPhys.83.863}, 
ions in traps\cite{RevModPhys.75.281}, 
optics\cite{Liu:2011aa}, and cold electric circuits\cite{Pekola2015aa} 
because well isolated systems can be prepared with these systems.
We, here,  point out that 
some molecules solved in isotropic liquid are well isolated 
and thus they can also be employed for 
studying open systems.

Our idea to employ molecules solved in isotropic liquid for studying an
open system comes from liquid-state Nuclear Magnetic Resonance 
(NMR) quantum computer experiments\cite{Cory1634,Gershenfeld350}. 
NMR quantum computing (NMR-QC) is one of the several proposals for realizing 
a quantum computer. 
The spins of nuclei in molecules solved in isotropic liquid are 
employed as qubits. NMR-QC is based on the fact that these spins have 
long coherence times and thus a lot of quantum controls can be applied to them
before losing their coherence. 
In other words, these are well isolated from their environment in the time
scale of NMR experiments. 
This isolation can be measured with the spin-lattice
relaxation times ($T_1$'s)\cite{Levitt2008}. 
They are characteristic times that the spins
(=~system) become thermal equilibrium because of interactions with their 
lattice (=~environment) and 
are often larger than 10~s while the times required for basic 
quantum operations, such as CNOT\cite{Nielsen2000}, 
are of the order of 1~ms. The inverses of the basic operation times
are the scales of interaction strengths between these spins. 

NMR-QC differs from other implementations of quantum computers 
in that it uses an ensemble of systems, or more than  $10^{15}$
molecules\cite{Cory1634,Gershenfeld350}, 
rather than a single system. 
This ensemble nature of NMR-QC makes difficult to implement a projection 
measurement although it is not impossible\cite{Kondo2016}.  Therefore,
the implementation of NMR-QC often requires some modifications from 
conventional quantum computing. 
Moreover, there are difficulties of scalability and pure-state preparation 
in NMR-QC\cite{Nielsen2000}. 
We believe, however, that these difficulties in NMR-QC are not very 
essential when we employ molecules in isotropic liquid for 
studying an open system. First, an ensemble average is 
necessary for obtaining statistical knowledge and thus the ensemble
nature of NMR can be a merit, rather than a demerit. Second, we are interested 
in studying experimentally a few-body system that is difficult to treat 
theoretically. Third, we employ molecules in isotropic liquid as a 
simulator of open systems and thus pseudo pure states\cite{Nielsen2000} 
are sufficient for our purpose.

\section{Theoretical Background}
\label{sec_theory}
In this section, we briefly review theoretical concepts and some
experiments about phenomena called relaxation from an intuitive 
view point.  
In order to understand relaxation, the 
concept of an open system is important and an open system 
may be experimentally realized with two different approaches. 
The first is adding a well controlled environment on an 
isolated system. The second is making a system of interest and 
its environment in an isolated system. In any way, we 
paradoxically need 
well, although approximately, isolated systems for these 
experiments. Therefore, we need to know some basic concepts 
of both open and isolated systems. 

At the end of this section, we discuss our experimental target,
an ensemble of molecules solved in isotropic liquid, 
as an ensemble of isolated systems. 

\subsection{Open System}
\label{ssec_closed_open_system}
A system of interest is interacting with its environment 
to a greater or less extent. Such a system is called to be {\it open}
because there are information flows between the system and its environment.
On the other hand, a system that is isolated from the rest is called 
a {\it closed} one. 

\subsubsection{Hamiltonian Description}
\label{h_description}
The {\it degree of openness} depends on the interaction strength between the
system of interest ($S$) and its environment ($E$) 
and on the time scale during which $S$ is influenced by $E$. 
The dynamics of the total system ($S$ and $E$) is 
determined by the total Hamiltonian
\begin{eqnarray}
\label{eq_open_s_H}
{\mathcal H} &=& {\mathcal H}_{\rm S} +  {\mathcal H}_{\rm I} 
+{\mathcal H}_{\rm E}, 
\end{eqnarray}
where ${\mathcal H}_{\rm S}, {\mathcal H}_{\rm I}, {\mathcal H}_{\rm E}$ are
the Hamiltonian of $S$ without the interactions with $E$, 
that of the interaction between $S$ and $E$, and 
that of $E$ without the interactions with $S$, respectively. 
The density matrix of the total system, $\rho$, 
is governed by the Liouville equation,
\begin{eqnarray}
\frac{d\rho}{dt} &=& - i [{\mathcal H}, \rho],
\end{eqnarray}
where we take the natural unit system so that $\hbar=1$. 
Its formal solution is
\begin{eqnarray}
\label{eq_open_total_dynamics}
\rho(t) &=& U(t) \rho(0) U^\dagger(t),
\end{eqnarray}
where $U(t) =e^{-i {\mathcal H}t}$ and the time parameter $t$ varies 
from 0 to $T$. $T$ may be finite or infinite.
We assume that ${\mathcal H}$ is time
independent. And then,
the density matrix of $S$ at $t$, $\rho_{\rm S}$(t), is given as 
\begin{eqnarray}
\label{eq_open_target_dynamics}
\rho_{\rm S}(t) &=& Tr_{\rm E} ( \rho(t)),
\end{eqnarray}
where $Tr_{\rm E}$ denotes the operation of
tracing out the environment freedom. 

$\rho_{\rm S}(t)$ is also 
described by the map $\Phi_t$.
\begin{eqnarray}
\rho_{\rm S}(0) &\rightarrow &  
\rho_{\rm S}(t) \stackrel{\rm def}{=}
\Phi_t \rho_{\rm S}(0) .
\end{eqnarray}
$\Phi_t$ must be {\it completely positive} and {\it trace preserving}
since $\Phi_t$ describes the time development of 
$\rho_{\rm S}$. Note that $\Phi_0 = I$ where $I$ is the identity map.  
Therefore, $\Phi_t$ is described by the operator sum
representation, as follows.
\begin{eqnarray}
\label{eq_open_target_kraus}
\Phi_t \rho_{\rm S}(0) 
= \sum_i \Omega_i(t) \rho(0) \Omega_i^\dagger (t),
\end{eqnarray}
where $\Omega_i$ is an operator and 
$\sum_i  \Omega_i(t) \Omega_i^\dagger (t)= I_{\rm S}$, where 
$I_{\rm S}$ is the identity matrix in the system Hilbert 
space\cite{RevModPhys.88.021002}.

In experiments, targets under study are intended to be as isolated, or
as decoupled, from their environments as possible so that they can be 
investigated without disturbances from environments. 
There are, however, always undesired interactions between them: 
these interactions make experiments complicated. 
Therefore, it is important 
to understand the open system characteristics. Moreover, 
if quantum mechanics is relevant in experiments, measurement devices are also 
considered to be a part of environment and thus understanding of the open system
characteristics is very essential. 

\subsubsection{Information Flow in Open System}
\label{ssec_relaxation}
A system of interest ($S$) interacts with its environment ($E$)
as discussed in 
\S~\ref{h_description} and finally becomes, or
{\it relaxes}, to a state in which the values of macroscopic 
quantities are stationary, universal with respect to different 
initial conditions, and predictable using statistical mechanics 
when the environmental degree of freedom is infinite.  
The final state must be a most probable ({\it typical} in the word 
of statistical mechanics) one.  
Independence of the final state on the initial one implies that 
$S$ loses the information of its
initial state. This {\it forgetting} mechanism is realized by 
transferring this information from $S$ to $E$, or entangling 
$S$ and $E$, by the interaction between them 
and then tracing out the freedom of $E$ as shown in 
Eq.~(\ref{eq_open_target_dynamics}). This can be interpreted as 
the information loss of $S$ by measuring with $E$\cite{Zurek_Phys_Today}.  

The final state that has perfectly 
lost the information of its initial state is, by definition of 
entropy,  a maximal entropy state. 
The information flow from $S$ to $E$ 
leads $S$  generically to thermalize towards the 
canonical mixed state\cite{Popescu2006}. This is called the General 
Canonical Principle. 
See also the references\cite{PhysRevLett.106.040401,PhysRevE.79.061103}. 
It was stressed that an ensemble-averaging operation 
is not essential. 

Let us consider a process towards a final state of $S$. 
Two qualitatively 
different processes exist\cite{RevModPhys.88.021002}.
 
\subsubsection{Markovian Relaxation}
Once the information of $S$ flows into $E$, it never
flows back to $S$. 
It is called a Markovian relaxation process. 

We assume that 
there is no correlation between $S$ and $E$ at $t=0$. 
A Markovian relaxation process can be approximately realized
when the degree of freedom of the environment is large enough and 
when $\tau_{\rm S}$ (the characteristic time of $S$) is much
longer than $\tau_{\rm E}$ (that of the environment).  

In the case of Markovian relaxation process, 
the map $\Phi_t$ defined
by Eq.~(\ref{eq_open_target_kraus}) satisfies 
\begin{eqnarray*}
\Phi_t \left( \Phi_{t'} \rho_{\rm S}(t'') \right)
&=& \Phi_{t+t'} \rho_{\rm S}(t'') 
\end{eqnarray*}
or in short $\Phi_t \Phi_{t'}= \Phi_{t+t'}$
for $t,t'\ge 0$. 
Therefore,  
\begin{eqnarray*}
 \Phi &=& \{ \Phi_t | 0 \le t \le T, \Phi_0 = I\}
\end{eqnarray*}
forms a semigroup\cite{QD_semiG_Appl}. 
The above equation is interpreted as 
\begin{eqnarray}
\label{eq_semi_g}
 \Phi_{t+ \Delta t} \, \rho_{\rm S}(0) &=& \rho_{\rm S}(t+\Delta t) 
=\Phi_{\Delta t}\left(  \Phi_t \, \rho_{\rm S}(0)\right) 
=\Phi_{\Delta t} \, \rho_{\rm S}(t).
\end{eqnarray}
Or, $\rho_{\rm S}(t+\Delta t)$ is determined only by 
$\rho_{\rm S}(t)$: the process is Markovian. 

The master equation determines the time evolution of 
$\rho_{\rm S}$ 
is in the Lindblad form\cite{0034-4885-77-9-094001}.
\begin{eqnarray}
\frac{d \rho_{\rm S}}{dt} &=& {\mathcal L}\rho_{\rm S}, \\
{\mathcal L} &=& -i [{\mathcal H}_{\rm S}, \rho_{\rm S}]
+ \sum_i \gamma_i \left( A_i \rho_{\rm S}(t) A_i^\dagger
-\frac{1}{2}\{ A_i^\dagger A_i, \rho_{\rm S}\}\right),
\end{eqnarray}
where $A_i$ is a time independent operator and $\gamma_i$ is a 
non-negative constant. A relaxation is caused by 
the dissipation, like friction, and stochastic 
terms\cite{Zurek_Phys_Today}. 

\subsubsection{non-Markovian Relaxation}
The information of $S$ flows into $E$ and sometimes 
flows back to $S$.  
It is called a non-Markovian relaxation process. 

This process is realized when there is correlation 
between $S$ and $E$ at $t=0$, or when 
$\tau_{\rm S} \gg \tau_{\rm E}$ is not satisfied. 
It can also happen when the degree of freedom of $E$  
is not large enough. Since we assume that the system of interest 
has the finite degree of freedom and 
thus the total system has the finite degree of freedom,
a recurrence in the total system dynamics is expected 
in a long term measurement according to the linearity of quantum mechanics. 

A non-Markovian relaxation process implies that $\Phi$ is not a semigroup.
In this case, the master equation determines the time evolution of 
$\rho_{\rm S}$ is very similar to the Lindblad form but 
different.
\begin{eqnarray}
\frac{d \rho_{\rm S}}{dt} &=& {\mathcal K}\rho_{\rm S}, \\
{\mathcal K} &=& -i [{\mathcal H}_{\rm S}(t), \rho_{\rm S}]
+ \sum_i \gamma_i (t)\left( A_i (t)\rho_{\rm S}(t) A_i(t)^\dagger
-\frac{1}{2}\{ A_i^\dagger(t) A_i(t), \rho_{\rm S}\}\right),
\end{eqnarray}
where ${\mathcal H}_{\rm S}(t)$ and $ A_i(t)$ are time-dependent operators and 
$\gamma_i(t)$ may become negative. Note that we assume that 
the inverse of $\Phi_t$, or $\Phi_t^{-1}$, exists although 
$\Phi_t^{-1}$ is not necessarily positive.  

\subsubsection{Divisibility and Measure of non-Markovianity}
Let us assume that $\Phi_t$ has an inverse $\Phi_t^{-1}$ for $t \ge 0$.
We define a two parameter map 
\begin{eqnarray}
 \label{eq_two_p_map}
 \Phi_{t,s} &\stackrel{\rm def}{=}& \Phi_t \Phi_s^{-1}, \hspace{2ex} t \ge s \ge 0,
\end{eqnarray} 
and $\Phi_{t,0}$ is defined by $\Phi_t$. Then, $\Phi_{t,0}$ and $\Phi_{s,0}$
can be related as 
\begin{eqnarray*}
 \Phi_{t,0} &=& \Phi_{t,s}\Phi_{s,0}.
\end{eqnarray*}
$\Phi_t$ is defined to be {\it P divisible} when 
$\Phi_{t,s}$ is positive. Similarly, it is {\it CP divisible}
when $\Phi_{t,s}$ is completely positive. 
If $\Phi_t$ is {\it P divisible}, $\Phi_{t,s}$ corresponds to
$\Phi_{\Delta t}$ in Eq.~(\ref{eq_semi_g}). And thus, 
{\it P divisible} $\Phi_t$ is Markovian. It is known 
that the reverse is true, too.  

We now define non-Markovianity of the map $\Phi_t$. 
Let us prepare an initial state $\rho_{\rm S}^1$ with a 
probability $p$ and another one $\rho_{\rm S}^2$ with a 
probability $1-p$. The operator $\Delta$ is defined as
\begin{eqnarray}
\Delta &\stackrel{def}{=}& p \, \rho_{\rm S}^1 - (1-p)\, 
\rho_{\rm S}^2.  
\end{eqnarray}
 $||\Delta||$, where $||*|| = Tr \sqrt{*^\dagger *}$, is
considered as the distinguishability of the two states. 
If $||\Phi_t \Delta||$ monotonically decreases with $t$ for all
$p$ and $\rho_{\rm S}^i$ combinations, $\Phi_t$ is said to be
Markovian. Then, the measure of non-Markovianity is defined as 
\begin{eqnarray}
 {\mathcal N}(\Phi) &\stackrel{def}{=}& 
\max_{p, \rho_{\rm S}^i} \int_{\sigma >0} \sigma dt,
\end{eqnarray}
where $\sigma (t) \stackrel{def}{=} \frac{d}{dt} ||\Phi_t \Delta ||$. 
If $\Phi_t$ is Markovian, $\sigma(t)$ is never positive and thus 
${\mathcal N}(\Phi) =0$. On the other hand, an information back flow 
causes an increase of $||\Phi_t \Delta ||$ and thus $\sigma(t)$ becomes 
positive in some time interval: ${\mathcal N}(\Phi) $ becomes positive.

\subsubsection{Collision Model}

The collision model\cite{PhysRev.129.1880} has been extended in order 
to model an open system 
that shows crossover from Markovian to non-Markovian relaxation 
according to its parameters\cite{PhysRevA.87.040103}. 
We,  here, discuss this extended model  because 
it is very simple yet powerful and because it was successfully 
applied to an atom in a dissipative cavity for a Lorentzian spectral 
density of bath modes\cite{PhysRevA.87.040103}.
Our experiments with molecules in isotropic 
liquids\cite{Kondo2016,Iwakura2017} may also be 
considered as realizations of variance of this extended collision model. 

The environment in the original collision model\cite{PhysRev.129.1880} 
is modeled as a large 
collection of non-interacting identical objects called ancillas. 
The system of interest, {\it S}, periodically interacts, or {\it collides}, with 
one of these ancillas. It is important that the collided ancilla will never
collide {\it S} again: this guarantees the one-way information flow from {\it S} 
to the environment. By controlling the strength and frequency of these
collisions, the relaxation rate of $\it S$ can be controlled. 

The ancilla interacts with {\it S} will disappear like in the original
collision model but is allowed to {\it collide}
another ancilla once, but only once, before disappearing 
in the extended collision 
model\cite{PhysRev.129.1880}.  This another ancilla is the one  
that will interact with {\it S} next: this collision between ancillas provides the
mechanism of memory in the environment. By controlling the collision 
between ancillas, the effect of memory can be controlled.

\subsubsection{Photon Passing through a Quartz Plate}
\label{ssec_engineered_open_system}

A photon is a well isolated system. It is ``too'' well-isolated that 
realization of a quantum computer with a photon is 
difficult\cite{Nielsen2000}. 
We, here, review a photon passing through a quartz plate\cite{Liu:2011aa}:
the polarization degree of freedom of photons $\ket{\lambda}$ 
($\lambda = H, V$ and $H= {\rm Horizontal}$ and $V= {\rm Vertical})$ 
was a system of interest, {\it S}, while its frequency degree of 
freedom $\ket{\omega}$ as its environment. 
The spectrum of the photons, the initial state of the environment, 
was controlled by passing them through a Fabry-P\'erot cavity. 
Their spectrum after the 
cavity depended on the incident angle to the 
cavity. Then, the interaction $U(t)$ between {\it S} and the environment 
was introduced by passing through a quartz plate as follows. 
\begin{eqnarray*}
U(t)\ket{\lambda} \otimes \ket{\omega} 
&=& e^{i n_\lambda \omega t}\ket{\lambda} \otimes \ket{\omega},
\end{eqnarray*}
where $n_\lambda$ is the polarization depend refraction 
index of the photons.  
The interaction strength was controlled by changing the interaction 
time $t$, or 
the thickness of the quartz plate where the photons 
passed. Then, full state tomography of the photons was carried 
out with a single photon detector.  Finally, the trace distance
and concurrence between the ``initial'' and final state were analyzed 
and these quantities were found to show non-monotonic decrease, 
or non-Markovian behavior\cite{RevModPhys.88.021002}. 
Note that the ``initial'' state was not
a real initial state because the single photon detector can only
provide a destructive measurement. One of an entangled photon 
pair was measured to obtain the ``initial'' state while the other 
photon of the pair was employed for the above experiment and 
its final state was measured. 

\subsection{Isolated system}

Dynamical chaos makes a system thermalize in classical 
statistical 
mechanics. An isolated quantum system, however, evolves linearly 
in time and the spectrum is discrete, and thus dynamical chaos itself
cannot occur. It has been a long question how an isolated quantum 
system relaxes to a thermal state. Some important concepts are reviewed
here\cite{Eisert:2015aa}. 

\subsubsection{Eigenstate Thermalization Hypothesis}

First of all, let us clarify the problem in relaxation of 
a quantum many-body system\cite{Rigol:2008aa}. 
An integrable quantum system of which 
dynamics is determined by a few constants called integrals cannot,
by definition, become thermalized because the initial state ``memory''
is kept in its dynamics. It is, however, that the condition of 
non-integrability is not sufficient. 

Let us consider a dynamics of a quantum many-body system of which 
initial state $\ket{\psi(0)}$ and its energy is $E_0$. 
It is expanded by the eigenvectors 
$\ket{\Psi_\alpha}$ of the system Hamiltonian ${\mathcal H}$ as, 
\begin{eqnarray*}
\ket{\psi(0)} &=& \sum_\alpha C_\alpha \ket{\Psi_\alpha},
\end{eqnarray*}
where $C_\alpha = \bra{\Psi_\alpha} \psi(0)\rangle$ and 
we assume that ${\mathcal H}$ is time independent and not degenerate
for simplicity.  
The time evolution of $\ket{\psi(t)}$ is given as
\begin{eqnarray*}
\ket{\psi(t)} &=& \sum_\alpha C_\alpha e^{-i {\mathcal H} t}\ket{\Psi_\alpha} 
=  \sum_\alpha C_\alpha e^{-i E_\alpha t}\ket{\Psi_\alpha} ,
\end{eqnarray*}
where $E_\alpha$ is an eigenvalue of $\ket{\Psi_\alpha}$. The expectation 
value of a few-body observable $O$ is given as,
\begin{eqnarray*}
\langle O \rangle &=& \bra{\Psi_\alpha(t)} O \ket{\Psi_\beta(t)}
= \sum_{\alpha, \beta} C_\alpha^* C_\beta e^{i(E_\alpha - E_\beta)t}
O_{\alpha, \beta}  ,
\end{eqnarray*}
where $O_{\alpha, \beta} = \bra{\Psi_\alpha} O \ket{\Psi_\beta}$. 
The long time average $\bar{O }$ must relax to 
$\displaystyle \sum_{\alpha} |C_\alpha|^2 O_{\alpha, \alpha}$. 
It means that this must also be the ensemble average 
of $O$ over the states of which energy is close to $E_0$,
 or a microcanonical average of $O$. 
\begin{eqnarray}
\label{eq_eth_base}
\sum_{\alpha} |C_\alpha|^2 O_{\alpha, \alpha}
& \approx & \bar{O }
\approx  \langle O \rangle_{\rm microcan}(E_0)
\stackrel{def}{=} 
\frac{1}{N_{E_0, \Delta E}}
\sum_{| E_\alpha -E_0 | < \Delta E} O_{\alpha, \alpha},
\end{eqnarray}
where $N_{E_0, \Delta E}$ is the number of eigenstates within the
energy window $[E_0 -\Delta E, E_0+\Delta E]$. 
The selection of $\Delta E$ is not important as long as a lot of 
states exist in this window. 

The problem is, now, clear: the left hand side of 
Eq.~(\ref{eq_eth_base}) depends on the initial 
state through $C_\alpha$ while there is no such dependence 
in the right hand side. In order to resolve this problem, 
it was conjectured that $O_{\alpha, \alpha}$ is approximately
constant for the eigenstate $\ket{\Psi_\alpha}$ of which 
eigenvalue satisfies 
$ | E_\alpha -E_0|<\Delta E$\cite{PhysRevA.43.2046,PhysRevE.50.888}.
This is called {\it eigenstate thermalization hypothesis} (ETH).
ETH is summarized as follows. 
\begin{quote}
 The initial state of energy $E_0$ is a linear combination of 
the (energy) eigenstates of which eigenvalues are near $E_0$.
These eigenstates have the property
of a thermal state but this property is not seen at first 
because of the coherence between them. However, this coherence 
disappears in time and the thermal state property
appears. This is the thermalization of a quantum many-body 
system. 
\end{quote}
Although ETH is still a hypothesis, it seems to be 
quite reasonable and promising\cite{Rigol:2008aa}.

\subsubsection{Generalized Gibbs Ensemble}

The term {\it equilibration} means that a system of interest  
undergoes relaxation to a state and stays there for a long time. 
An equilibrated state may depend on its initial state: this is different 
from thermalization. A certain system may become, or equilibrate to, 
a quasistationary state before thermalized.  
This phenomenon is called 
{\it prethermalization}\cite{PhysRevB.84.054304,Gring1318}. 
Such a quasistationary state is thought to be described 
as an generalized Gibbs ensemble (GGE) state that 
is a  maximum entropy state with some constraints 
that related with the initial state, or conserved quantities = 
{\it integral}s. 

It is worth noting that the term {\it integrable} is taken 
differently by researchers. In the reference\cite{Rigol:2008aa}, 
it was stated that a quantum {\it integrable} system had 
conserved quantities of which
number was much more than one but less than the dimension of its 
Hilbert space. On the other hand, this term was employed more restrictively
like in the case of classical mechanics: a quantum {\it integrable}
system had the same number of conserved quantities as that of the
degrees of the freedom\cite{PhysRevLett.106.040401}. 
Depending on the usage, GGE systems are said to be either 
integrable or non-integrable. 

Regardless of the usage of the term {\it integrable}, the GGE system 
does not thermalize because its integrals, or the information of 
the initial state, exist at $t \gg 1$. 

\subsubsection{Many-Body Localization}

Many-body localization also prevents a system from 
thermalization\cite{doi:10.1146/annurev-conmatphys-031214-014726}.
It is analogous to the well known Anderson 
localization\cite{PhysRev.109.1492} that  
a disordered potential make a single-particle 
wave function localize in real space. In contrast, the many-body 
localization occurs as a quasi particle localization in Fock space 
because of a random potential and interactions among components 
of the many-body system.  
A quasi particle localization means that the information of 
the initial state is also kept.

\subsubsection{Scrambling}

A concept recently attracting attention in connection with
thermalization in quantum information and gravity fields is 
{\it scrambling}. 
Let us consider a quantum dynamics where initial states that are 
very similar but orthogonal with each other evolve to be quite different. 
Such a chaotic behavior is referred to as {\it scrambling}\cite{Shenker2014}
and often called the quantum butterfly effect by comparing with 
its classical counter part. 
The scrambling was originally discussed with a black hole:
what happens for the information possessed by a falling object into a black hole 
when it crosses the event horizon? The information of the falling object is 
accessible before crossing but it is not after. Therefore, it is conjectured 
that a black hole is Nature's fastest scrambler\cite{1126-6708-2008-10-065}. 

\subsubsection{Lieb-Robinson Bound}
The Lieb-Robinson bound\cite{L-R1972} implies that the speed of 
information propagation 
in a non-equilibrium system has a certain limit. Such a propagation is also 
called the {\it light cone}-like information flow because it is similar to 
the propagation of light\cite{Eisert:2015aa} in the theory of relativity. 
It is formally described as follows. Let us consider two observables, 
$O_1$ and $O_2$. 
\begin{eqnarray}
\label{eq_L-R_bound}
\| [O_1(t), O_2]\| \le c \|O_1\|\, \|O_2 \| {\rm min}\{|O_1|,|O_2|\}
e^{-\mu(d(O_1,O_2)-v |t|)},
\end{eqnarray}
where $\|\cdot\|$ is the operator norm, $ d(O_1,O_2)$ the distance 
between the support of the observables, 
$|O_1|$ and $|O_2|$ the size of their supports, $v\ge 0$ is the velocity,
and $c$ and $\mu$ are positive constants.  

The constant velocity propagation 
of the two-point parity correlation (information) was reported in the 
experiment with a one-dimensional quantum gas in an optical 
lattice\cite{Cheneau2012}. 

\subsubsection{Out of Time Order Correlation (OTOC)}
A special type of correlation functions, called out-of-time-order 
correlation (OTOC) function $F(\tau)$,  
has recently been attracting attention in order to quantify information, or 
entropy, flow and to probe scrambling of information\cite{PhysRevA.94.040302}. 
It is thus related to the Lieb-Robinson
bound\cite{PhysRevLett.117.091602}, too. 
Let us consider two unitary operators $W$ and $V$ that 
are commuting at time $\tau =0$.
Starting from $[W_\tau, V]=0$ at $\tau=0$, let us consider the change of 
$[W_\tau,V]$ in time $\tau$ and  assume that it becomes non-zero. 
Here, $W_\tau= U_\tau^\dagger W U_\tau$ and $U_\tau =e^{-i{\mathcal H}\tau}$.  
It implies that $W$ that is initially commuting with $V$ becomes 
non-commuting after $\tau$ because of the interactions generated 
by ${\mathcal H}$.

The average of the square of 
$[W_\tau,V]$ is given as, 
\begin{eqnarray}
\label{oq_OTOC}
\langle | [W_\tau, V]|^2\rangle &=& 2 (1-\Re[F(\tau)]) \\
{\rm where} && F(\tau) \stackrel{def}{=} 
\langle W^\dagger_\tau V^\dagger W_\tau V\rangle . 
\end{eqnarray}
$F(\tau)$ is the out-of-time-order correlation function of $W$ and $V$. 
In order to clarify the physical meaning of $F(\tau)$ and thus 
$\langle | [W_\tau, V]|^2\rangle$, $F(\tau)$ is re-written as 
\begin{eqnarray*}
F(\tau) &=& \langle O_{\rm B}^\dagger(V,W,\tau) \, 
O_{\rm F}(V,W,\tau)\rangle , \\
{\rm where} &&  \, O_{\rm F}(V,W,\tau) \stackrel{def}{=} W U_\tau V, \\
&& O_{\rm B}(V,W,\tau) \stackrel{def}{=} U_\tau V U_\tau^\dagger W U_\tau. 
\end{eqnarray*}

When $W$ and $V$ are properly selected, $F(\tau)$ has been proved 
to be equivalent to the 
second  R\'{e}nyi entropy\cite{PhysRevB.77.064426,FAN2017707}. 
The Lyapunov exponent
$\lambda_\mathrm{L}$  can also be defined with 
$1-F(\tau)\propto e^{\lambda_\mathrm{L}\tau}$ 
when $W$ and $V$ are hermitian. 
The OTOC in many-body systems can be employed in order to distinguish a many-body
 localized phase from an Anderson localized one while a normal correlation 
 function cannot\cite{FAN2017707}.   

Although the OTOC function, $F(\tau)$, is very useful, 
its measurement is not trivial for a many-body quantum system 
because of the time reversal operation $U_\tau^\dagger$ 
appeared in $O_{\rm B}(V, W, \tau)$.
Despite of the difficulty, some OTOC's, however, have been measured 
in a NMR\cite{PhysRevX.7.031011} and trapped ion\cite{Garttner2017}
experiments.

\subsubsection{The Second Law of Thermodynamics in Closed System}

Equilibrium thermodynamics is based on the fact that the system 
of interest is contact with the canonical bath, or environment. 
The second law means that the system evolves quasistatically 
in such a way that its entropy increases. The increase of entropy 
indicates the  information flow from the system to the environment.  

The second law of thermodynamics, and also the fluctuation theorem, 
in a closed and pure system have been discussed\cite{PhysRevLett.119.100601}.  
This closed system consists of the system of interest, {\it S}, 
and its environment. It was pointed out that the second law of 
thermodynamics is applicable for {\it S} 
with some error even if the environment, that is a part of the closed system,  
is not a canonical bath, but if it is an energy eigenstate and the weak ETH 
is satisfied. Here, ``weak'' means 
``less restrictive''\cite{PhysRevLett.119.100601}. 
This conclusion is based on the fact that 
{\it S} cannot distinguish a weak ETH environment from the canonical bath 
because the speed of the information flow (the Lieb-Robinson 
bound\cite{L-R1972}) is limited and thus  enough information 
for distinguishing them cannot 
be collected within a limited time.

\subsection{molecules solved in isotropic liquid as isolated systems}
\label{subsec_molecule}
\begin{figure}[b]
\centerline{
\includegraphics[width=4.7cm]{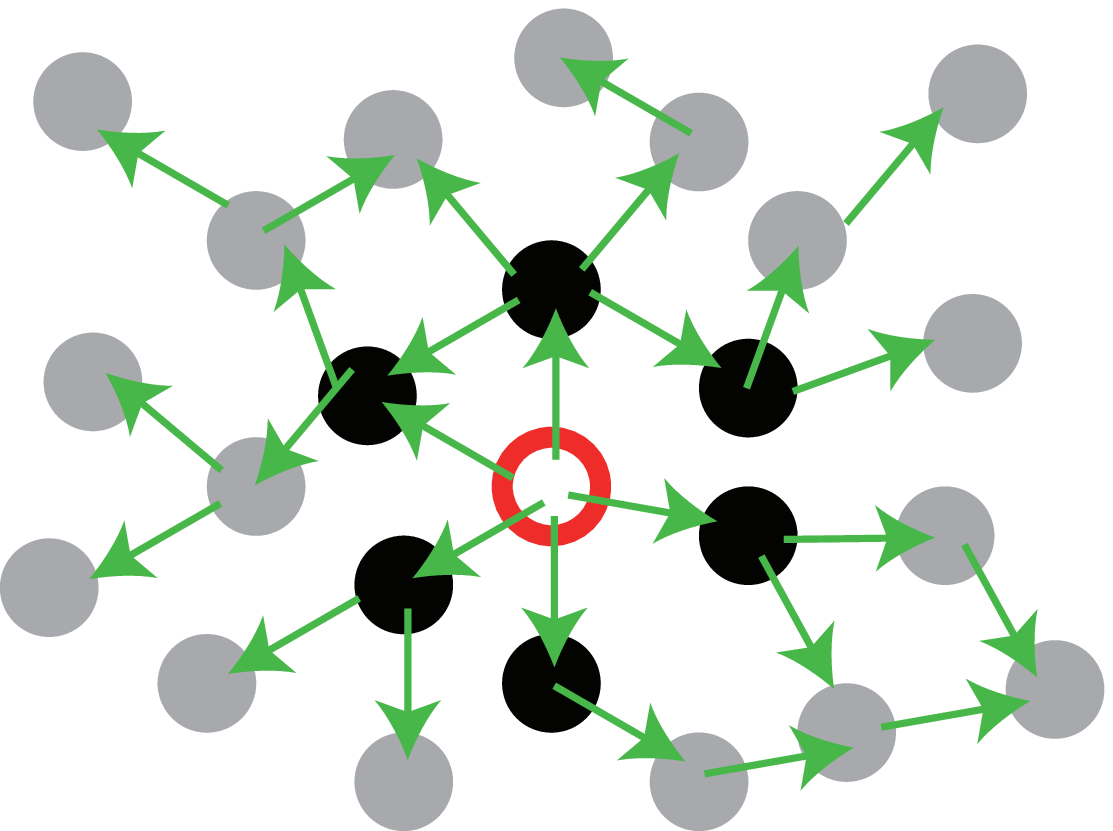}
\includegraphics[width=4.7cm]{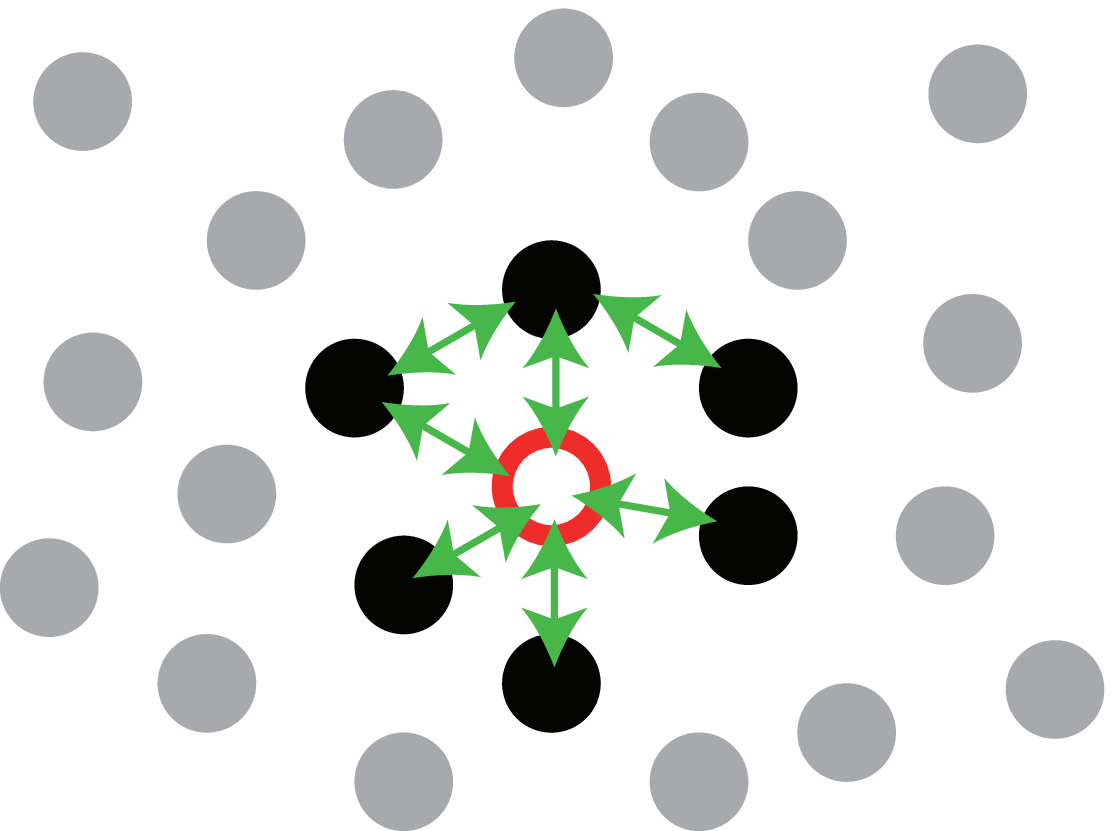}
}
\centerline{
(a)\hspace{30ex}(b)
}
\vspace*{8pt}
\caption{ Molecules in liquids. The arrows indicate the flow of information 
originally in the red open circle spin, or interactions among nuclear spins,  
in the molecules and liquids. The red and black spins consist of 
the molecule, while the gray ones are spins in liquids. (a) Open system, 
or the information flows into the liquid. (b) Closed system, or the
information stays inside of the molecule. 
\label{fig_open_close_system}}
\end{figure}

Molecules solved in liquid are under influence  
of solvent molecules. In principle, the nuclear spins in
the molecule consist of an open system, as shown in 
Fig.~\ref{fig_open_close_system}~(a). 
We, however, have to take into account that molecules 
in isotropic liquid 
\begin{itemize}
\item are under an external strong magnetic field, and 
\item are rapidly moving. 
\end{itemize}
The strong external magnetic field dominates the spin dynamics and 
thus simplifies it. Or, the original Hamiltonian which determines 
the spin dynamics may be replaced by a more simpler one: secular 
approximation\cite{Levitt2008}. On the other hand, the rapid molecular 
motion causes interactions to fluctuate in time. The fluctuating interaction may be 
replaced by its time averaged one: motional averaging\cite{Levitt2008}. 

The secular approximation and motional averaging make 
effective interactions much simpler than the original ones 
in molecules solved in isotropic liquids\cite{Levitt2008}.   
\begin{itemize}
 \item Intramolecular interactions are described as 
\begin{eqnarray}
\label{eq_mol_ham}
 {\mathcal H} &=& \sum_j \omega_{0, j} \frac{\sigma_{z,j}}{2}
 + \sum_{j<k} J_{j,k} \frac{\vec{\sigma}_{j} \cdot \vec{\sigma}_{k}}{4}
\end{eqnarray}
where 
$$ \vec{\sigma}_i = 
\overbrace{I \otimes I \otimes \cdots \otimes I}^{1\sim i-1}
\otimes \vec{\sigma}
\otimes \overbrace{I \otimes I \otimes \cdots \otimes I}^{i+1\sim n}, 
\hspace{2ex} {\rm and} \hspace{2ex} 
\vec{\sigma} = (\sigma_x, \sigma_y, \sigma_z). $$
$\sigma_k$ is a standard Pauli matrix, 
$\omega_{0,j}$ the isotropic chemically shifted Larmor frequency of 
the $j$'th spin, and $J_{j,k}$ the interaction strength between 
the $j$'th and $k$'th spins. 
$J_{j,k}$ is often measured in Hz in NMR textbooks, but
it is measured in \mbox{rad/s} as $\omega_{0,j}$ here. 
 \item Interactions among spins inside and outside molecules are 
dipole-dipole interactions. 
\end{itemize}
If the dipole-dipole interactions 
among spins inside and outside molecules 
could be ignored, the spins 
in the molecules are regarded as an isolated spin
system\cite{Levitt2008},  
as shown in Fig.~\ref{fig_open_close_system}~(b). 
We also note that the strength of the intermolecular dipole-dipole
interaction can be evaluated by measuring $T_1$ because this flips the
spin in the molecule and causes the spin-lattice relaxation. 

Therefore, molecules in isotropic liquids investigated 
with NMR may be approximately isolated 
from the environment, their Hamiltonian is simple, and 
they are able to be measured with NMR techniques.

\section{Experiments}
\label{sec_exp}
We have been working on molecules in isotropic liquids with NMR: 
we have employed them in order 
to study ideas of quantum computation\cite{PhysRevA.70.052319},  
robust quantum controls\cite{doi:10.1143/JPSJ.80.054002,Ichikawa4671}, 
Bang-Bang controls\cite{Iwakura2017}, and so on. Our NMR spectrometer is 
a standard one for precise chemical analysis, or JEOL ECA-500.

We, here, present experiments in which molecules in isotropic liquid  
are regarded as an ensemble of isolated, at least approximately, 
systems. 
An ensemble of cold atoms is regarded 
as a well isolated system\cite{RevModPhys.83.863,Eisert:2015aa} 
and the isolated system has often been theoretically studied 
with the cold atoms in mind. In this section
we point out that an ensemble of molecules in isotropic liquid, which is 
much easier to treat, is also a good model to study an isolated 
system. We note that molecules in isotropic liquid can simulate 
quantum registers in quantum computation and 
that their dynamics is approximately unitary, at least in a 
short time. In other words, some molecules in isotropic liquids are 
widely accepted as approximately isolated 
systems\cite{Cory1634,Gershenfeld350}. 

We recommend a reader who is not familiar with NMR to refer 
a standard NMR textbook, such as one by Levitt\cite{Levitt2008}.
We also provided a crush course of NMR 
for NMR quantum computation\cite{Kondo2006,Kondo2009} which may be 
a convenient start point.   
The concept of our experiments is discussed 
in \S~\ref{subsec_exp_concept}.  
The state, or information, 
of a part of this  molecule is characterized by its transverse 
magnetization and its dynamics 
after an abrupt change is tracked with NMR techniques. 
Then, our experiments are summarized in \S~\ref{sub_ext_cm} 
and \ref{sub_SE_Isolated_M}.

\subsection{Relaxation Experiments with Molecules in Isotropic Liquids} 
\label{subsec_exp_concept}
We designed two types of experiments: {\it realization of the extended 
collision model} and {\it a system and environment in an isolated molecule}. 

\subsubsection{Density Matrix and Free Induction Decay Signal}
We performed two kinds of relaxation experiments.
The information that we tracked in our experiments was 
the transverse magnetization of one of the nuclear spins 
in the molecule,
perpendicular to the applied strong static magnetic field. 
From the quantum mechanical view, the thermally equilibrium 
density matrix $\rho_{\rm th}$ of 
a molecule in isotropic liquid at room temperature $T$,
see the Hamiltonian (\ref{eq_mol_ham}),  is well approximated as
\begin{eqnarray}
\label{eq_d_m_th}
\rho_{\rm th} &=& \left( \frac{I}{2}\right)^{\otimes n} 
+ \sum_j \frac{  \omega_{0,j}}{2 k_B T} \frac{\sigma_{z,j}}{2^n}
\end{eqnarray}
because the thermal energy $k_B T$ is much larger than that of 
magnetic ones $  \omega_{0,j}$ and because the energies 
associated with the interactions $  J_{i,j}$ are 
even much smaller than the magnetic 
ones\cite{Levitt2008,Kondo2006,Kondo2009}. 
Let us assume that we measure 
the first spin and then $\rho_{\rm th}$ can be rewritten as 
\begin{eqnarray*}
\label{eq_d_m_th}
\rho_{\rm th} &=& 
\left( \frac{I}{2}\right)^{\otimes n} 
\left(1- \frac{  \omega_{0,1}}{2 k_B T}\right)
+ \frac{  \omega_{0,1}}{2 k_B T} 
\left(
\begin{array}{cc}
 1 & 0 \\
 0 & 0
\end{array}
\right) \otimes 
\frac{I^{\otimes (n-1)}}{2^{n-1}}
+ \sum_{j \ne 1} \frac{  \omega_{0,j}}{2 k_B T} 
\frac{\sigma_{z,j}}{2^n}.
\end{eqnarray*}
Note that $(I/2)^{\otimes n}$ is not visible in NMR because 
${\rm Tr}(\vec{\sigma}_j (I/2)^{\otimes n}) = \vec{0}$. 
Then, we apply the operation 
$R(\pi/2, \pi/2)\otimes I^{\otimes (n-1)}$ on $\rho_{\rm th}$, 
where 
\begin{eqnarray}
\label{eq_rf_pulse}
R(\beta, \phi) &\stackrel{def}{=}& 
e^{- i\beta(\cos \phi \sigma_x + \sin \phi \sigma_y)/2 }.
\end{eqnarray}
This rotates the first spin magnetization 
by the angle $\beta$  around the axis 
in the $xy$-plane with an angle $\phi$ from the $x$-axis. 
This operation can be realized by applying a rotating, 
or equivalently oscillating, magnetic 
field of which frequency is $\omega_{0,1}$ 
for a certain period\cite{Levitt2008,Kondo2006,Kondo2009}. 
We obtain the following initial state 
\begin{eqnarray}
\rho(0) &=& 
\frac{1}{2}
\left(
\begin{array}{cc}
 1 & 1 \\
 1 & 1
\end{array}
\right) \otimes 
\frac{I^{\otimes (n-1)}}{2^{n-1}}
\end{eqnarray}
by re-normalizing 
$\displaystyle \frac{  \omega_{0,j}}{2 k_B T} \rightarrow 1$. 
We neglect $(I/2)^{\otimes n}$ and the other spin terms 
since they are not observable.

Let us consider the dynamics in the frame where spins are  
seen in their own rotating frame of which frequencies are 
their Larmor ones. In this frame, Hamiltonian~(\ref{eq_mol_ham})
is reduced to 
\begin{eqnarray}
\label{eq_mol_ham_rf}
 {\mathcal H} &=&
 \sum_{j<k} J_{j,k} \frac{\sigma_{z,j} \, \sigma_{z,k}}{4},
\end{eqnarray}
if we assume that $|\omega_{0,j} - \omega_{0,k}| \gg |J_{j,k}|$
for our molecules. This assumption is 
called the weak coupling limit\cite{Levitt2008}. 

 We observe the relaxation process, or more precisely 
dephasing one, of the spin~1  under 
the Hamiltonian~(\ref{eq_mol_ham_rf}). 
When the freedom of the other spins are traced out, the density
matrix of the spin~1, $\rho_1(t)$, can be described as 
\begin{eqnarray*}
\rho_1(t) &=& n_x(t) \frac{\sigma_x}{2} +n_y(t) \frac{\sigma_y}{2}
+\frac{I}{2}. 
\end{eqnarray*}
Note that the diagonal terms are constant under 
Hamiltonian (\ref{eq_mol_ham_rf}). 
A complex Free Induction Decay (FID) signal $S(t)$ is given as 
\begin{eqnarray}
\label{eq_FID}
S(t) = n_x(t)+i\, n_y(t) &=& 
Tr\left(\left(\sigma_x+i\, \sigma_y \right) \,\rho_1(t) \right). 
\end{eqnarray} 
Thus, $n_x(t)$ and $n_y(t)$ are called the real and imaginary part of 
the complex FID signal, respectively. For example, 
$\displaystyle S(0) = 1$, while the perfectly dephased state 
$ \displaystyle \rho_1(\infty) =  \frac{1}{2} I 
$ 
gives $S(\infty) = 0$. The trace distance between $\rho_1(t)$ 
and $\rho_1(\infty)$, $D(\rho_1(t), \rho_1(\infty))$, is given as 
\begin{eqnarray}
\label{eq_Tr_d}
D\left(\rho_1(t), \rho_1(\infty) \right) &\stackrel{def}{=}&
\frac{1}{2}Tr\left( 
\sqrt{\left(\rho_1(t) -\rho_1(\infty)\right)^\dagger
\left(\rho_1(t) -\rho_1(\infty)\right)}\right)
\nonumber \\
&=&\frac{1}{2}\sqrt{n_x^2(t)+n_y^2(t)},
\end{eqnarray}
and thus $S(t)$ is a good measure of the trace distance in our 
experiments where we observe the dephasing.

\subsubsection{Decoupling}
There is a very interesting technique
in NMR called decoupling\cite{Levitt2008}. The interaction 
strengths inside molecules in isotropic liquids are usually 
of the order of 100~Hz or less in frequency unit. Therefore, if 
one can flip-flop the spins much faster than these 
frequencies, for example by applying rf pulses on them, 
these spins are nullified in average. We employed 
this technique in order to control the 
number of effective (active) spins in the molecule. 

\subsubsection{Realization of the Extended Collision Model}

The extended collision model\cite{PhysRevA.87.040103} may be 
re-formulated as follows. The system of interest, {\it S}, frequently 
collides one ancilla. If this ancilla is reset before next collision 
with {\it S}, the information flow is one-way and thus {\it S} shows 
a Markovian relaxation. A large collection of ancillas is not 
necessary. If the reset of the ancilla is not perfect, or if the 
interval $\tau_{\rm reset}$ between resetting the ancilla is large 
so that the information backflow occurs, the ancilla provides 
the mechanism of memory of the environment. {\it S} may show a 
non-Markovian relaxation. 

\begin{figure}[t]
\centerline{
\vspace{-3ex}
\includegraphics[width=4cm]{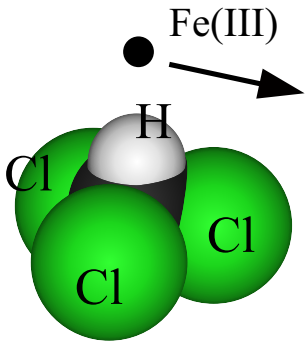}
}
\vspace*{8pt}
\caption{Realization of the extended collision model with a chloroform 
molecule and Fe(III) magnetic impurity. The $^{13}$C nuclear 
spin is {\it S} while the  hydrogen one 
corresponds to the ancilla. Note that chlorine nuclear spins are 
magnetically inert. The resetting mechanism is provided by 
randomly moving Fe(III) paramagnetic impurities. 
\label{fig_CHCl3}}
\end{figure}

Our reported experiments 
with chloroform molecules\cite{Kondo2016,Iwakura2017} may be 
considered as a faithful realization of the above re-formulated 
extended collision model. The $^{13}$C nuclear spin is {\it S} while 
the  hydrogen one 
corresponds to the ancilla. Note that chlorine nuclear spins are 
magnetically inert. The resetting mechanism is provided by 
randomly moving Fe(III) paramagnetic impurities. 
See Fig.~\ref{fig_CHCl3}.  
Because of $r^{-3}$ dependence of the dipole field generated by
the Fe(III) magnetic impurities, where $r$ is the distance from it, 
the $^{13}$C nuclear spin 
is approximately not influenced by the 
Fe(III) impurities compared with the hydrogen spin. 
$\tau_{\rm reset}$ 
is of the order of its longitudinal relaxation time $T_{\rm 1, H}$.
The Hamiltonian that determines the spin dynamics of the chloroform 
molecule is  
$\displaystyle {\mathcal H} = J \frac{\sigma_z \otimes \sigma_z}{4}$,
where $J =  2 \pi \cdot 215$~rad/s, and thus 
the time scale of the spin dynamics is of the order of 
$\displaystyle \frac{2 \pi}{J}$. See also Eq.~(\ref{eq_mol_ham_rf}). 
Therefore, the frequentness of the resetting the ancilla is 
measured with $\displaystyle \frac{2 \pi}{J \, T_{\rm 1, H}}$. 
More rigorous treatment with the operator sum 
representation\cite{Nielsen2000,Kraus1971,PhysRevA.57.4153} 
can be found in our previous publications\cite{Kondo2016,Iwakura2017}. 

 In this work, we show the experiments with tetramethylsilane (TMS)
molecules with a small $J$ in order to demonstrate non-Markovian to 
Markovian crossover of relaxation. The small $J$ means that  
$\displaystyle \frac{2 \pi}{J \, T_{\rm 1,H}}$ can be large 
with the same $T_{\rm 1, H}$.

\subsubsection{System and Environment in an Isolated Molecule}

We also design other experiments by employing molecules in isotropic 
liquids as an ensemble of isolated systems.  
An interacting nuclear spin system in one isolated molecule is divided 
into two parts. One of them is treated as a system of interest.
We are inspired by the experiments with ultra cold atoms that form 
a well isolated system,  but we are interested in the behavior
of a finite system.  
We are interested in the minimum size of the left that is 
able to act as environment of the first.

We employed $^{13}$C enriched transcrotonic-acid and normal
4,4-dimethyl-4-silapentane-1-sulfonic acid (DSS) molecules.  
DSS is very popular as a frequency
standard in NMR experiments. 

\subsection{Realization of the Extended Collision Model}
\label{sub_ext_cm} 

\subsubsection{Less Frequent Resetting}
\begin{figure}[b]
\centerline{
\includegraphics[width=3cm]{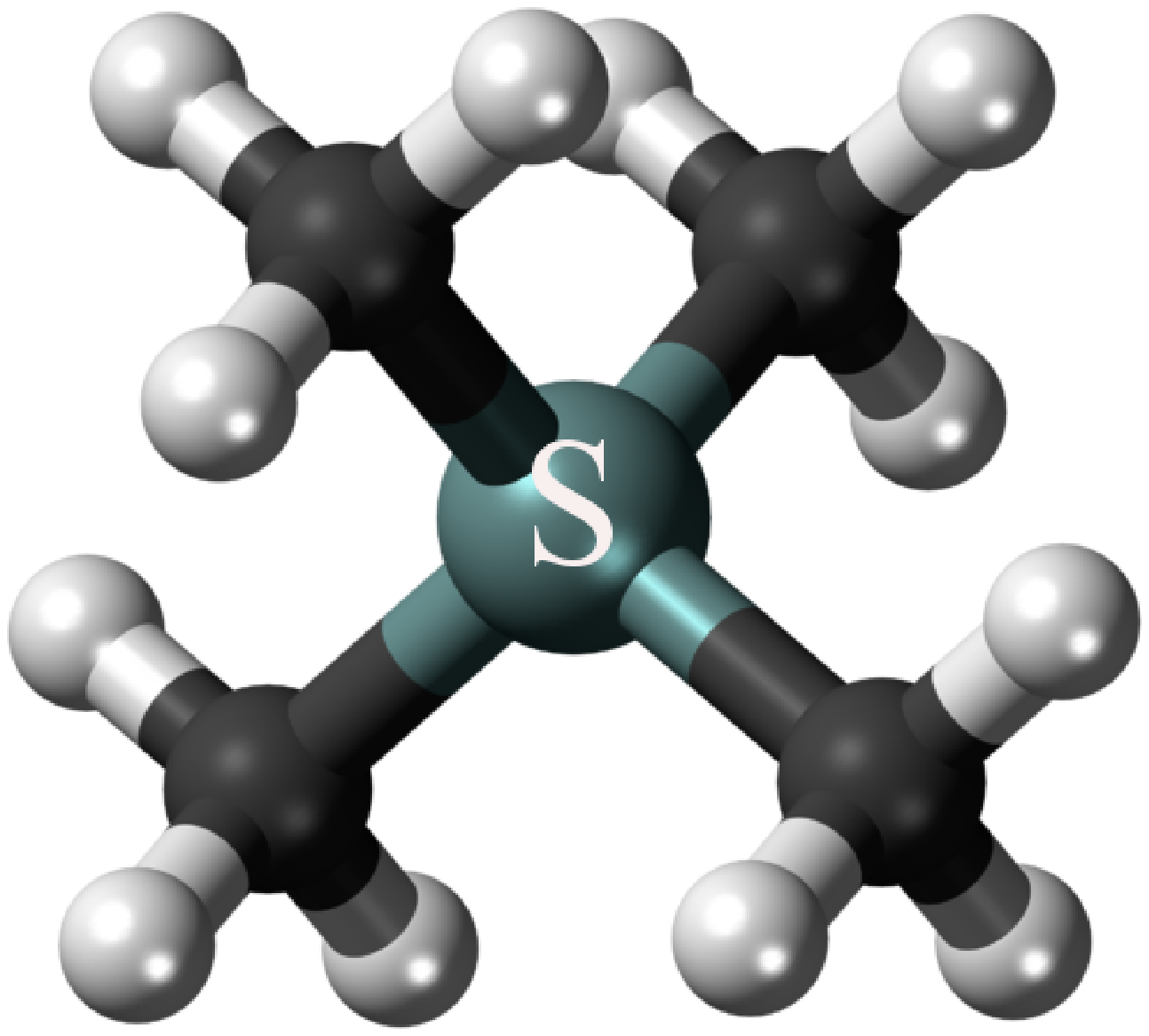} 
\hspace{10ex}
\includegraphics[width=3cm]{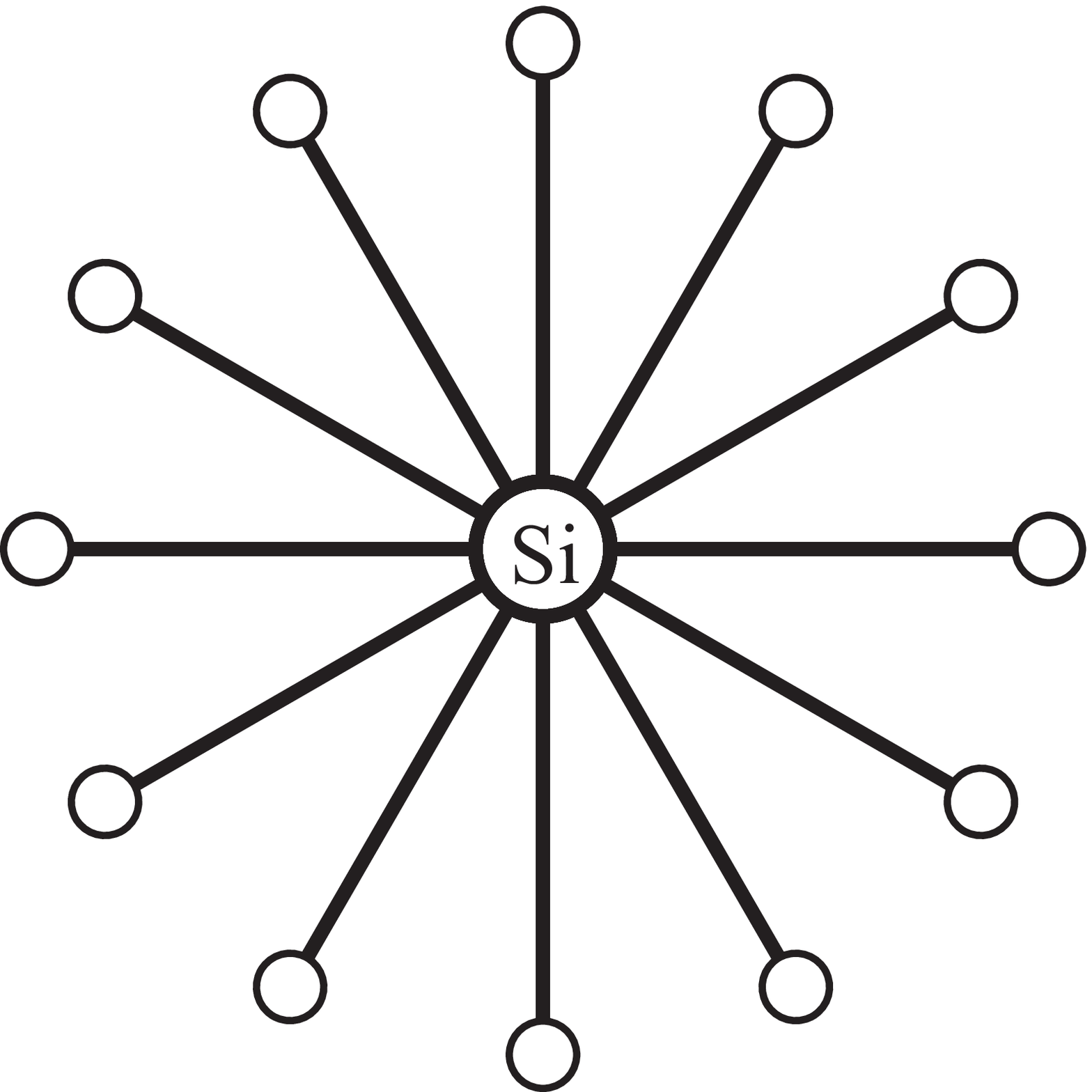} 
}
\centerline{(a)\hspace{28ex}(b)}
\vspace*{8pt}
\caption{Tetramethylsilane (TMS). (a) Structure. The center sphere marked S
is a silicon, the four black ones are carbon, and the twelve white
ones are hydrogen atoms, respectively. (b) Interaction network. 
The silicon spin equally interacts with the twelve hydrogen spins 
represented with small circles because of the molecular structure symmetry.
The carbon spins can be ignored because they are not $^{13}$C enriched.  
\label{fig_tms}}
\end{figure}

We employed tetramethylsilane (TMS) of which molecular structure is 
shown in Fig.~\ref{fig_tms}~(a). This molecule is often employed for 
the frequency standard of $^{13}$C and hydrogen and is very popular 
in standard NMR experiments. The natural abundance of the silicon isotope 
$^{29}$Si of which spin is 1/2 and active in NMR is about 5~\% and thus 
these spins are easily measured with NMR. We assigned 
the center silicon nuclear spin ($^{29}$Si) as the system of interest 
{\it S}, while the twelve hydrogen spins are ancillas.  
Our sample is not $^{13}$C enriched and thus we can ignore 
the carbon spins because its natural abundance of $^{13}$C is only 1~\%. 
The interaction network is shown in Fig.~\ref{fig_tms}~(b).  It is 
very simple because the molecular structure is very symmetric.
The interaction strength is $2 \pi \cdot 6.6$~rad/s which is much smaller 
than that of chloroform. 
There are no interactions among hydrogen spins because of the symmetry.

The sample is a mixture of 2.89~g normal 
(neither $^{13}$C nor $^{29}$Si is enriched) tetramethylsilane 
and 2.53~g acetone-d6 (deuterized acetone). 
$T_1$ of silicon spin is 16~s and that of 
hydrogen is 10~s. Therefore, this molecule is well isolated 
from the solvent in the time scale of few tenths of second.

\begin{figure}[t]
\centerline{
\includegraphics[width=8cm]{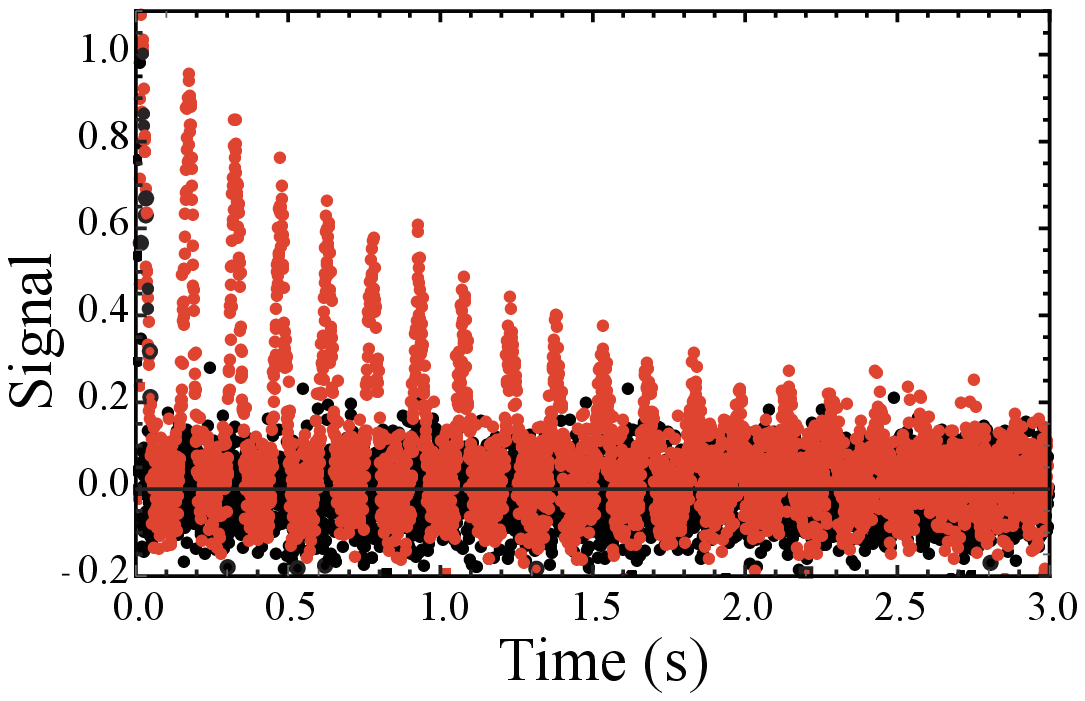}
\hspace{-10ex}
\includegraphics[width=8cm]{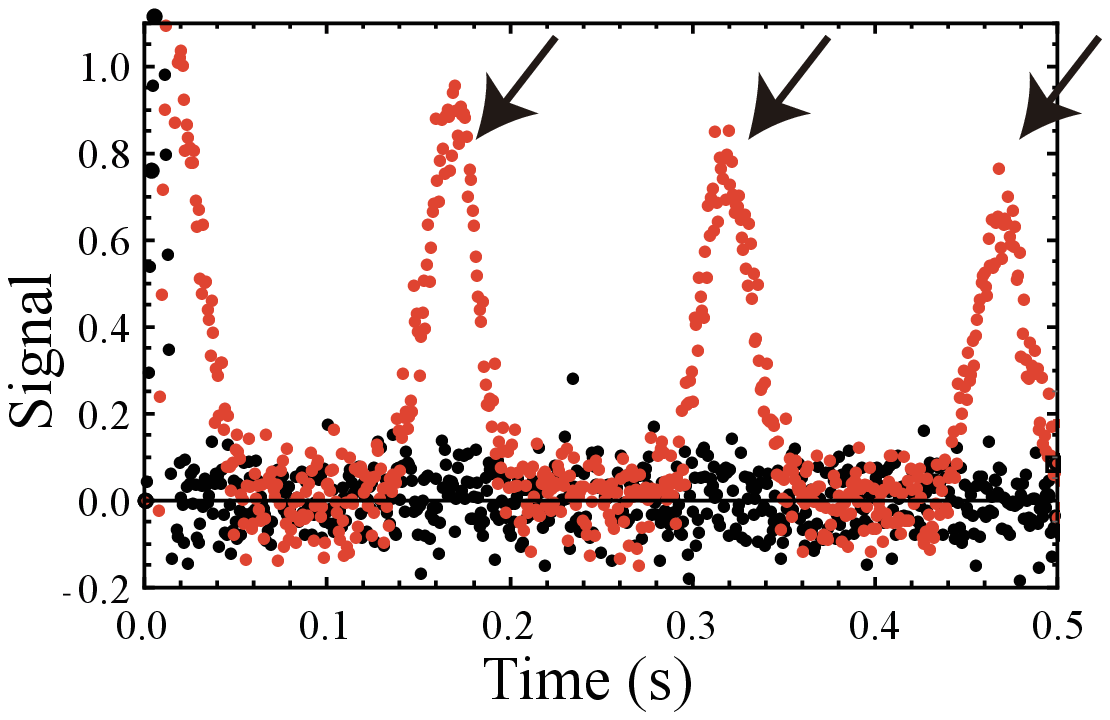}
}
\centerline{\hspace{10ex}(a)\hspace{40ex}(b)}
\vspace*{8pt}
\caption{ FID signal of silicon spin in tetramethylsilane molecule. The red 
 (black) points are the real (imaginary) part of the complex FID signal. 
(a) More than ten times revivals are seen.
(b) The first 0.5~s of 
the FID signal 
is shown. 
\label{fig_tms1}}
\end{figure}

The FID signal of the central silicon nuclear spin is 
shown in Fig.~\ref{fig_tms1}. 
The FID signal is a good measure of the trace distance 
$D(\rho_1(t), \rho_1(\infty))$ as discussed with 
Eqs.~(\ref{eq_FID}, \ref{eq_Tr_d}).
The FID signal shows revivals which indicate that the information originally
at the silicon spin moves to the surrounding hydrogen spins and then flows back
to the silicon spin again and again. More than ten revivals are seen. 

This experimental result can be interpreted as follows:
The hydrogen spins (= ancillas) are not frequently reset 
(large $T_{\rm 1, H}$) and thus 
the environment well {\it remembers} the history of the interaction 
with the system of interest.

\subsubsection{Frequent Resetting}

We introduced the resetting mechanism by adding some magnetic impurities, 
as we discussed. 
We prepared two samples with magnetic impurities of
19~mM and 40~mM. Their $T_1$'s of silicon spins are  
2.6~s (19~mM) and 1.4~s (40~mM). Therefore, the direct influence 
of the magnetic impurity can approximately be ignored in the first 
few tenths of a second. On the other hand,  $T_{\rm 1, H}$'s
are 140~ms (19~mM) and 70~ms (40~mM) and are much shorter than 
the time scale determined by the interaction strength 
$2 \pi \cdot 6.6$~rad/s: it implies that 
resetting the ancillas is very frequent. 
We also conclude that the magnetic impurities 
independently reset ancillas because 
\mbox{$(T_1 \times {\rm impurity \hspace{1ex} concentration})$}
is constant\cite{Iwakura2017}. 

\begin{figure}[htb]
\centerline{
\includegraphics[width=8cm]{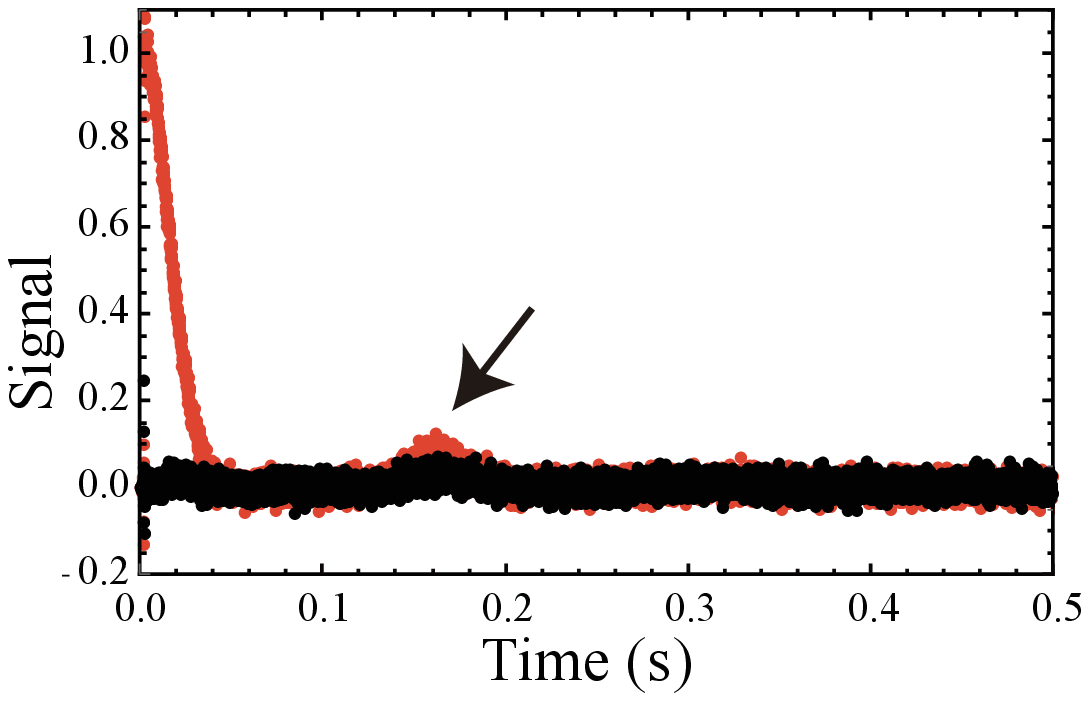}
\hspace{-10ex}
\includegraphics[width=8cm]{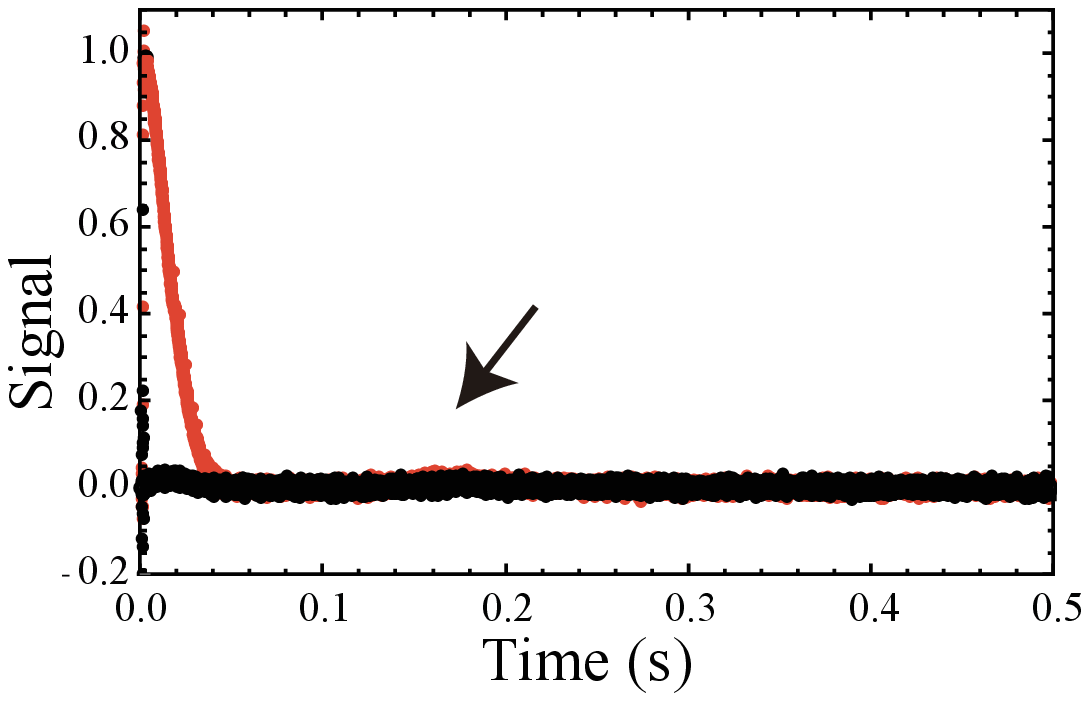}
}
\centerline{\hspace{10ex}(a)\hspace{40ex}(b)}
\vspace*{8pt}
\caption{FID signals with Fe(III) impurities of 
(a) 19~mM and (b) 40~mM. 
The crossover from non-Markovian to Markovian relaxation 
is observed in increasing the concentration of Fe(III), or 
the frequency of resetting.  See, also Fig.~\ref{fig_tms1}~(b) 
for comparison. 
\label{fig_tms5}}
\end{figure}

Revivals are not visible with the 40~mM sample, while 
the small revival is seen with the 19~mM one, as shown 
in Fig.~\ref{fig_tms5}. 
It can be interpreted as follows: resetting ancillas is frequent 
enough so that the information flow becomes one-way, or the system
of interest in the 40~mM sample shows a Markovian relaxation. 
We observe the crossover from non-Markovian to Markovian relaxation
by increasing the concentration of the magnetic impurities.

\subsection{System and Environment in an Isolated Molecule}
\label{sub_SE_Isolated_M}

\subsubsection{Molecules with large degrees of freedom}
\begin{figure}[b]
\centerline{
\includegraphics[width=10cm]{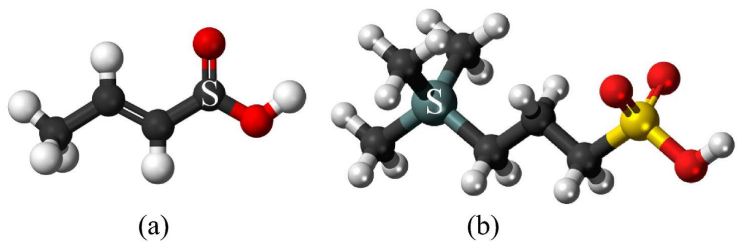}
}
\vspace*{8pt}
\caption{Molecular structures of (a) transcrotonic-acid and 
(b) normal 4,4-dimethyl-4-silapentane-1-sulfonic acid. 
The spheres marked S are the systems of interest.
S in the transcrotonic acid molecule is a $^{13}$C spin 
in the carboxy group, 
while that in the DSS molecule a silicone spin. 
The black spheres are carbon, the white ones are hydrogen, 
the red ones are oxygen, and the yellow one is sulfa, respectively. 
The numbers of interacting spins are nine (fifteen) in the
transcrotonic acid (DSS). \label{fig_TCA_DSS}}
\end{figure}

We employed $^{13}$C-enriched transcrotonic-acid and normal 
4,4-dimethyl-4-silapentane-1-sulfonic acid
(DSS) molecules solved in heavy water as an approximately 
isolated molecules with large degrees of freedom. 
Their molecular structures are shown in Fig.~\ref{fig_TCA_DSS}.  
The number of spins in  $^{13}$C-enriched transcrotonic-acid is nine, while 
that of normal DSS fifteen. Note that the natural abundance of $^{16}$O 
without spin is almost 100~\% and that the hydrogens 
in the carboxy and sulfo group
in Fig.~\ref{fig_TCA_DSS} are detached 
in water. 

The $^{13}$C-enriched transcrotonic-acid molecule
has often been employed in NMR quantum computer experiments\cite{Knill2000aa}.  
In other words, this molecule 
has been accepted to be a well isolated spin system. 
What is interesting with the DSS molecule is similarity in its
molecular structure with that of TMS. TMS is usually employed in NMR
experiments as a frequency standard in organic solvents, while DSS 
in water. When one of the hydrogen atoms in TMS is replaced with a carbon
chain which has a sulfo group at the end, DSS is obtained. This carbon 
chain 
makes DSS solvable in water.  
On the other hand, this modification in molecular structure increases
the degree of freedom in our experimental point of view
since the molecular structure of DSS is less symmetric than that of TMS. 
It is noteworthy that the degree of freedom of the TMS molecule is 
not so large compared with $^{13}$C-enriched transcrotonic-acid or normal 
DSS because of the symmetry of TMS molecule although the number of spins 
in it is as large as thirteen. 

$T_1$'s of $^{13}$C in the transcrotonic acid solved in D$_2$O 
(about 70~mM) were 8~s (methyl group), 
10~s, 10~s, and 14~s (carboxy group) from left to right 
in Fig.~\ref{fig_TCA_DSS}~(a), while those
of the hydrogen spins were 4~s (methyl group), 6~s, and 6~s. 
On the other hand, $T_1$'s of $^{29}$Si in DSS (0.083~g solved in
0.74~g of D$_2$O) were 7~s, while those
of the hydrogen spins were 3~s (methyl group) and 1.5~s 
(the others) in  Fig.~\ref{fig_TCA_DSS}~(b).
We concluded that  both the transcrotonic acid and DSS molecules  
are well isolated 
in the first few tenths of a second of FID signals.

\subsubsection{DSS}
\label{subsub_DSS}
\begin{figure}[htb]
\centerline{
\includegraphics[width=13cm]{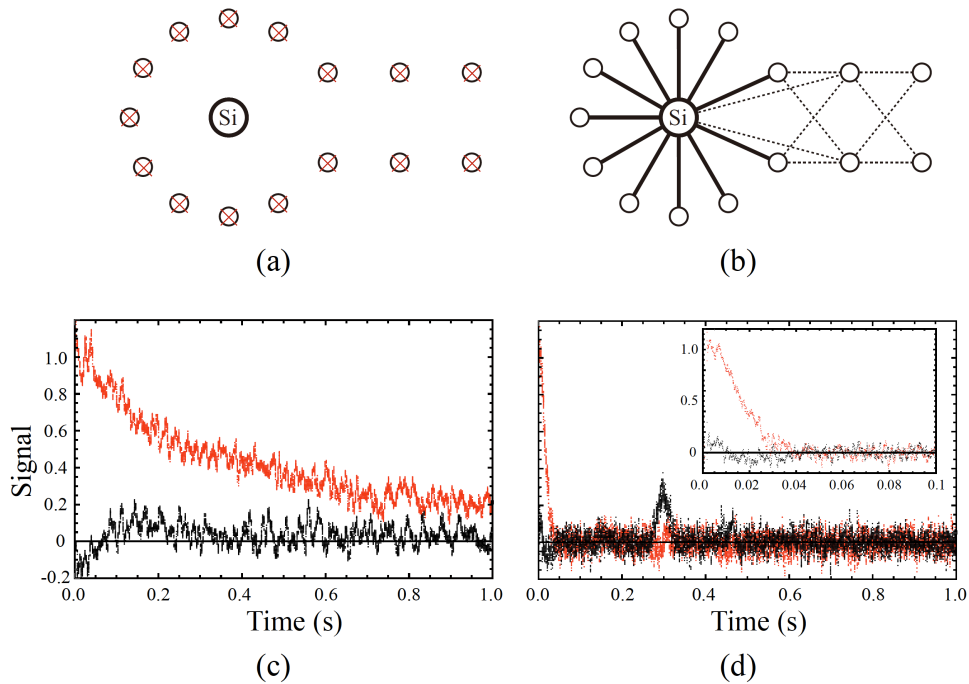}
}
\vspace*{8pt}
\caption{ 
Interaction networks (a, b) and FID signals (c, d) of DSS in D$_2$O. 
The FID signals of the silicon spin in DSS were measured. 
(a, c) All hydrogen spins are nullified. 
(b, d) All spins are active and thus the number of degrees of freedom is  
large. (d) The inset shows the first 0.1~s of the FID signal. 
The thick solid (thin dotted) lines in (b) represent interactions 
through one (two) carbon atom(s). The small circles represent hydrogen 
spins. 
\label{fig_DSS}}
\end{figure}

We performed experiments with the DSS molecule, shown 
in Fig.~\ref{fig_TCA_DSS} (b), as that with 
the large degree of freedom. 
Figure~\ref{fig_DSS}~(a, b) show 
the interaction networks with and without hydrogen decoupling. 
There is a very large difference between them. 
Figure~\ref{fig_DSS}~(c) shows the FID signal of 
the silicon spin 
when all hydrogen spins were decoupled. In this case, the number of the 
spins in the molecule was effectively one, only the silicon spin. 
The signal decay indicates that the molecule was not perfectly isolated 
from the environment. One can, however, state that it was approximately 
isolated in a short time scale of 0.1~s.  

The FID signal of the silicon spin is shown in Fig.~\ref{fig_DSS}~(d) 
when all hydrogen spins were active. This FID signal is very different 
from that with decoupling (Fig.~\ref{fig_DSS}~(c)) or that of TMS 
(Fig.~\ref{fig_tms1}). If we stopped the measurement of the FID signal 
at 0.1~s,
we might conclude that the silicon spin relaxes perfectly. In other words, 
the finite system of the 15 hydrogen spins act as environment for the silicon 
spin and induces relaxation of the silicon spin within 0.1~s.  
On the other hand, we can clearly see a revival at 0.3~s although it is 
very small compared with those observed in the FID signal of TMS. 
We interpret that the information flows into the carbon chain, reflected 
at its end, and flows back to the silicon nuclear spin. In other words, 
we observe the finite speed of the information flow, or the Lieb-Robinson
bound\cite{L-R1972}. The interaction strengths between the hydrogen nuclear 
spins in 
the carbon chain, measured from the spectra of the hydrogen spins,  are 
of the order of $2\pi \cdot 10$~rad/s and  consistent with
the revival at 0.3~s.  

\subsubsection{Transcrotonic Acid}
\label{subsub_TCA}

We employed the transcrotonic acid molecule, 
shown in Fig.~\ref{fig_TCA_DSS} (a), in order to 
demonstrate the influence of the degree of freedom on 
the information flow. 

\begin{figure}[htb]
\centerline{
\includegraphics[width=10cm]{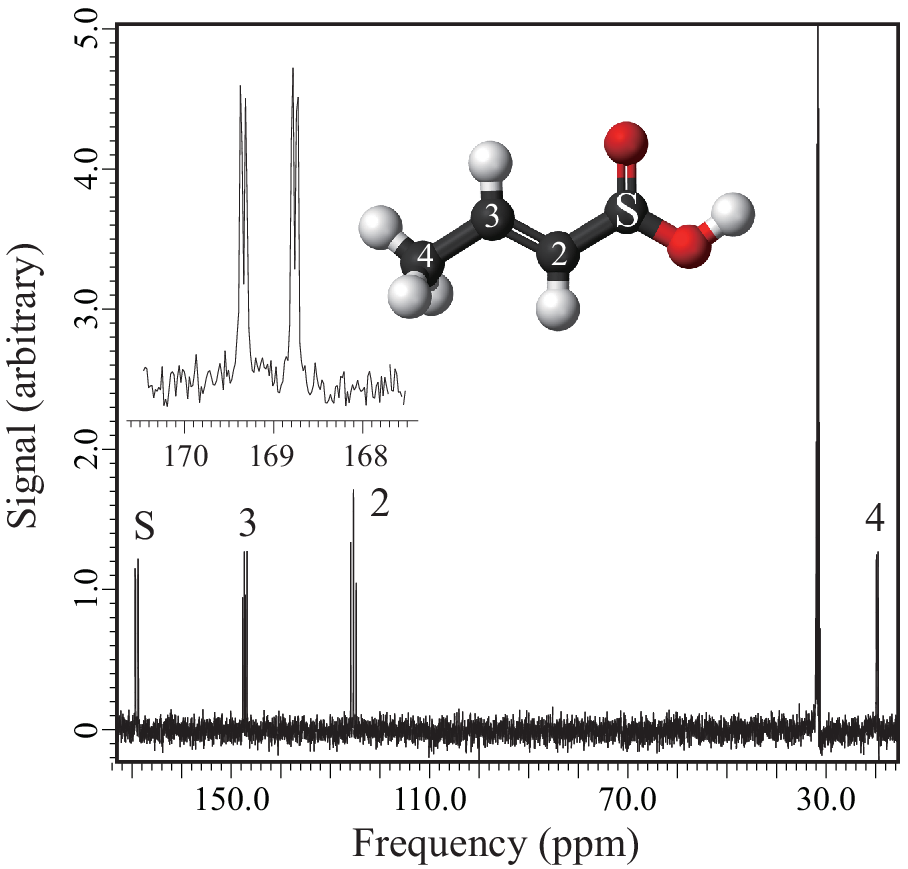}
}
\vspace*{8pt}
\caption{
Spectrum of transcrotonic acid in D$_2$O after a short rf-pulse that
rotates all $^{13}$C spins 
into the $xy$-plane when all hydrogen spins were decoupled. 
The peaks and the $^{13}$C spins in the molecular model are 
corresponded by number. The peak at 33~ppm may be originated from 
some impurity. 
The detailed structure of the peak marked S is shown in the inset. 
See the text for more details. 
\label{fig_TCA_sp}}
\end{figure}

Figure~\ref{fig_TCA_sp} shows the $^{13}$C spectrum obtained 
after operating $R(\pi/2, \pi/2)^{\otimes 4}$ on $\rho_{\rm th}$ by 
applying a short rf-pulse on all $^{13}$C spins 
and when all hydrogen spins were decoupled. Close view of the peak 
marked S, see the inset of Fig.~\ref{fig_TCA_sp}, reveals 
that it consists of two groups of peaks and each
group consists of two peaks: four peaks in total. These four peaks
correspond to four different states of No.~2 and 3 $^{13}$C spins, 
or $\ket{00}, \ket{01}, \ket{10}$, and $\ket{11}$. 

We employed another NMR technique called a soft pulse\cite{Levitt2008}.  
A Gaussian envelope pulse with as long as 40~ms duration can 
turn the only magnetization at 169.4~ppm in Fig.~\ref{fig_TCA_sp}. 
The state that we obtained was  
\begin{eqnarray*}
\rho_{\rm ini} &=& 
\frac{1}{2}
\left(
\begin{array}{cc}
 1 & 1 \\
 1 & 1
\end{array}
\right) \otimes 
\left(
\begin{array}{cc}
 0 & 0 \\
 0 & 1
\end{array}
\right) \otimes 
\frac{I^{\otimes (n-2)}}{2^{n-2}},
\end{eqnarray*}
where $n$ is the total number of active spins.  
If the hydrogen spins were decoupled, $n=4$. On the other hand, 
$n = 9$ if the hydrogen spins were active. 
The first spin is the $^{13}$C in the carboxy group. 
Then, the information (transverse magnetization) flows  
into the spin chain. 

\begin{figure}[htb]
\centerline{
\includegraphics[width=13cm]{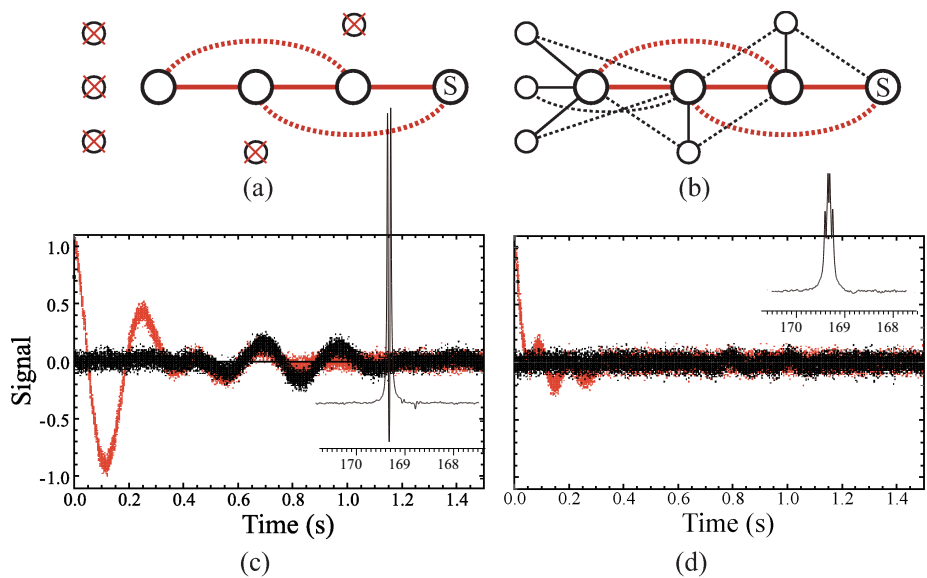}
}
\vspace*{8pt}
\caption{
Interaction networks (a, b) and FID signals (c, d) of 
transcrotonic acid in D$_2$O. The large circles represent $^{13}$C 
nuclear spins,
while the small ones hydrogen nuclear spins. The red thick 
(black thin) lines represent interactions between $^{13}$C$-^{13}$C 
($^{13}$C$-$hydrogen). The solid (dotted) lines represent the nearest 
(next nearest) interactions. 
The FID signals of the $^{13}$C nuclear spin 
(marked S) in the carboxy group were measured. (a, c) All the hydrogen
nuclear spins are decoupled, or the degree of freedom is small. 
The large revivals are seen. 
(b, d) All the spins are active and thus the degree of freedom is 
large. The revivals are smaller. 
The spectra correspond to these FID signals are also shown in the insets. 
Note that there is no peak at 168.8~ppm. 
\label{fig_TCA}}
\end{figure}

Figure~\ref{fig_TCA}~(a, b) show the interaction networks. 
The interaction network with the hydrogen spins (b) is much 
more complicated than that without the hydrogen spins (a). 
It implies that the degree of freedom of (b) is
much larger than that of (a). 
The FID signals of the $^{13}$C nuclear spin in the carboxy group 
are shown in Fig.~\ref{fig_TCA}~(c, d) with and without
hydrogen decoupling, respectively. 
Large revivals are seen in Fig.~\ref{fig_TCA}~(c), while 
they are much smaller in Fig.~\ref{fig_TCA}~(d). 
The more degrees of freedom, the smaller revivals. 

\section{Conclusions}
We first very briefly discussed the basic ideas related 
with relaxation phenomena 
that should be a good start point for a person who is interested 
in them. We, however, recommend serious 
readers to refer, at least, the papers listed here and 
references therein. We also apologize to the readers 
for our not-comprehensive reference list. 
We then provided our experimental results with molecules 
in isotropic liquids as one unique experimental 
approach to  investigating 
relaxation phenomena. We employed these molecules as 
isolated few body systems.

We believe 
that our approach may be interesting and effective 
because there are various molecules available and because
experiments can be done with a standard commercial NMR
equipment. Only the idea is important and no difficult experimental
techniques are required. 
These experiments must complement those 
with ultra cold atoms\cite{RevModPhys.83.863}, 
ions in traps\cite{RevModPhys.75.281}, 
optics\cite{Liu:2011aa}, and cold electric circuits\cite{Pekola2015aa}.  

\section*{acknowledgments}
We would like to thank Hayato Nakano for participating in fruitful discussions. 
YK would like to thank a partial support of JST CREST
Grant Number JPMJCR1774. 

\bibliographystyle{ws-ijmpcs}
\bibliography{bib}

\end{document}